\documentclass[preprint,showpacs,preprintnumbers,superscriptaddress,amsmath,amssymb]{revtex4}

\usepackage{color}
\usepackage{graphicx}
\usepackage{dcolumn}
\usepackage{bm}
\usepackage{amsmath}
\usepackage{nicefrac}

\RequirePackage{xspace}

\newcommand{\ie}{\textit{i.e.}\xspace}
\newcommand{\eg}{\textit{e.g.}\xspace}

\newcommand{\cf}{\textit{cf.}\xspace}

\newcommand{\etc}{\textit{etc.}\xspace}

\newcommand{\units}[1]{\ensuremath{\,\mathrm{#1}}}

\newcommand{\fracd}[3][]{\displaystyle\frac{d^{#1}#2}{d#3^{#1}}}

\RequirePackage{xspace}
\newcommand{\YBCO}[1][]{\ensuremath{\mathrm{Y Ba_2 Cu_3 O_{7#1}}}\xspace}

\newcommand{\sgn}{\mathop{\mathrm{sgn}}\nolimits}

\def\stt#1{%
\if#1.
\let\next=\relax
\else
  \if#1u
    \upharpoonleft
  \else\if#1d
    \downharpoonleft
  \else\if#1U
    \upharpoonright
  \else\if#1D
    \downharpoonright
  \else\if#1F
    \Uparrow
  \else\if#1A
    \Downarrow
  \else\if#1s
    \uparrow
  \else\if#1a
    \downarrow
  \else\if#1r
    \rightarrow
  \else\if#1l
    \leftarrow
  \else\if#1R
    \Rightarrow
  \else\if#1L
    \Leftarrow
  \else\if#1>
    \longrightarrow
  \else\if#1<
    \longleftarrow
  \else\if#1
  \relax
  \else
  ?\message{Unknown symbol #1 in the argument of macro state}
  \fi\fi\fi\fi\fi\fi\fi\fi\fi\fi\fi\fi\fi\fi\fi
\let\next=\stt
\fi
\next
}

\newcommand{\ZeroPi}{\ensuremath{0\text{-}\pi}\xspace}
\newcommand{\opi}{\ensuremath{0\text{-}\pi\text{-}}\xspace}
\newcommand{\OPI}[1][20]{\ensuremath{#1\times(\opi)}\xspace}

\newcommand{\img}{\ensuremath{\langle j_s \rangle}}

\begin{document}

\title{Visualizing supercurrents in ferromagnetic Josephson junctions with various arrangements of  $0$ and $\pi$ segments}

\author{C.~G\"urlich}
\author{S.~Scharinger}
\affiliation{%
  Physikalisches Institut -- Experimentalphysik II and Center for Collective Quantum Phenomena,
  Universit\"at T\"ubingen, Auf der Morgenstelle 14,
  D-72076, T\"ubingen, Germany
}

\author{M.~Weides}
\altaffiliation[Current address: ]{%
  Department of Physics, University of California,
  Santa Barbara, CA 93106, USA
}

\affiliation{%
  Institute of Solid State Research and JARA-Fundamentals of Future Information Technology,
  Research Center J\"{u}lich,
  D-52425 J\"{u}lich, Germany
}

\author{H.~Kohlstedt}
\affiliation{%
  Nanoelektronik, Technische Fakult\"{a}t, Christian-Albrechts-Universit\"{a}t zu
Kiel, D-24143 Kiel, Germany
}

\author{R.G.~Mints}
\affiliation{%
  The Raymond and Beverly Sackler School of Physics and Astronomy, Tel
Aviv University, Tel Aviv 69978, Israel
}

\author{E.~Goldobin}
\author{D.~Koelle}
\author{R.~Kleiner}
\affiliation{%
  Physikalisches Institut -- Experimentalphysik II and Center for Collective Quantum Phenomena,
  Universit\"at T\"ubingen, Auf der Morgenstelle 14,
  D-72076, T\"ubingen, Germany
}

\date{\today}

\begin{abstract}
      Josephson junctions with ferromagnetic barrier can have positive or negative critical current depending on the thickness $d_F$ of the ferromagnetic layer. Accordingly, the Josephson phase in the ground state is equal to 0 (a conventional or 0 junction) or to $\pi$ ($\pi$ junction).
When $0$ and $\pi$ segments are joined to form a ``\ZeroPi junction'', spontaneous supercurrents around the \ZeroPi boundary can appear. Here we report on the visualization of supercurrents in superconductor-insulator-ferromagnet-superconductor (SIFS) junctions by low-temperature scanning electron microscopy (LTSEM). We discuss data for rectangular  0, $\pi$, \ZeroPi, $0$-$\pi$-$0$ and  \OPI[20] junctions, disk-shaped junctions where the \ZeroPi boundary forms a ring, and an annular junction with two \ZeroPi boundaries. Within each 0 or $\pi$ segment the critical current density is fairly homogeneous, as indicated both by measurements of the magnetic field dependence of the critical current and by LTSEM. The $\pi$ parts have critical current densities $j_c^\pi$ up to $35\units{A/cm^2}$ at $T = 4.2\units{K}$, which is a record value for SIFS junctions with a NiCu F-layer so far. We also demonstrate that SIFS technology is capable to produce Josephson devices with a unique topology of the \ZeroPi boundary.
\end{abstract}



\maketitle

\section{Introduction}
\label{sec:intro}

As predicted more than 30 years ago\cite{Bulaevskii77}, Josephson junctions can have a phase drop of $\pi$ in the ground state. Such $\pi$ junctions are now intensively investigated, as they have a great potential for applications in a broad range of devices ranging from classical digital circuits\cite{Terzioglu97,Terzioglu98,Ustinov03a,Ortlepp06} to quantum bits\cite{Ioffe99,Blatter01,Yamashita05,Yamashita06}. Nowadays, $\pi$ Josephson junctions can be fabricated by various technologies, including junctions with a ferromagnetic barrier \cite{Ryazanov01,Kontos02,Blum02,Bauer04,Sellier04,Oboznov06,Weides06b,Vavra06,Bannykh09}, quantum dot junctions\cite{vanDam06,Cleuziou06,Jorgensen07} and nonequilibrium superconductor-normal metal-superconductor Josephson junctions \cite{Baselmans99, Baselmans02, Huang02}

In the simplest case the supercurrent density $j_s$ across the junctions is given by the first Josephson relation
\begin{equation}
  j_s = j_c \sin\phi,
\label{Eq:CPR}
\end{equation}
with the critical current density $j_c > 0$ for a 0 junction and $j_c < 0$ for a $\pi$ junction. Here, $\phi$ is the gauge invariant phase difference of the superconducting wave function across the junction (Josephson phase).

Particularly superconductor-insulator-ferromagnet-superconductor (SIFS) junctions \cite{Kontos02,Weides06b,Bannykh09} are promising since, in contrast to other types of $\pi$ junctions, they exhibit only small damping at low temperatures, which is necessary to study Josephson vortex dynamics as well as to use them as active elements in macroscopic quantum circuits.

Now consider a junction in the $x$--$y$ plane, which has a region with critical current density $j_c^0>0$ (0 region) and another region having $j_c^\pi<0$ ($\pi$ region). For the sake of simplicity let us assume that the boundary between $0$ and $\pi$ regions runs along the $y$ direction. When $\phi$ is different from $0$ or $\pi$ the supercurrents flow in opposite directions on the two sides of the 0-$\pi$ boundary, forming a vortex, with its axis coinciding with the 0-$\pi$ boundary (along the $y$ direction), that carries a magnetic flux $\Phi=\pm\Phi_0/2$  ($\Phi_0\approx2.07\times10^{-15}\units{Wb}$ is the flux quantum) \cite{Bulaevskii78, Xu95,Goldobin02a}. This is true if the junction length $L$ in $x$ direction is much larger than the Josephson penetration depth
\begin{equation}
  \lambda_J = \sqrt{\frac{\Phi_0}{2\pi |j_c| \mu_0 d'}}
  . \label{Eq:lambda_J}
\end{equation}
Here $\mu_0 d'$ is the inductance per square (with respect to in-plane currents) of the superconducting electrodes forming the junction. For junctions having electrode thicknesses larger than the London penetration depth $\lambda_L$, $d' \approx 2\lambda_L$. Experimentally, such semifluxons have first been studied in the context of cuprate grain boundary junctions\cite{Kirtley96,Kirtley99} or zigzag ramp junctions between Nb and  \YBCO \cite{Hilgenkamp03}. Here, the sign change of the $d$-wave order parameter of the cuprates leads to the formation of 0-$\pi$ facets. In junctions with a ferromagnetic barrier the value (and the sign) of the critical current density crucially depends on the thickness $d_F$ of the F-layer\cite{Kontos02,Weides06b}. A junction consisting of various 0 and $\pi$ segments can, thus, be formed by selectively etching the F-layer to produce two thicknesses $d_F^0$ and $d_F^\pi$ of the F-layer such that they correspond to critical current densities $j_c^0$ and $j_c^\pi$ with opposite signs and $j_c^0\approx|j_c^\pi|$\cite{Weides06a}.

In the cuprate/Nb zigzag junctions \cite{Smilde02,Hilgenkamp03,Ariando05,Guerlich09} the facets should be oriented along the crystallographic $a$ and $b$ axes of the cuprate electrode, imposing certain topological limitations to the 0-$\pi$ boundary. In contrast, the SIFS technology allows almost any 2D shape of the 0-$\pi$ boundary and therefore offers a higher degree of design flexibility. Below, we show an example where this boundary forms a loop. Even intersecting 0-$\pi$ boundaries should be feasible, \eg, by arranging 0 and $\pi$ regions in a checkerboard pattern. Unfortunately, the present SIFS technology based on a NiCu ferromagnetic layer produces a maximum $|j_c^\pi|$ which is much lower than $j_c\sim1\units{kA/cm^2}$ of standard Josephson tunnel junctions. Although $j_c^\pi$ at $T = 4.2\units{K}$ has been increased from some $\units{mA/cm^2}$ for the first junctions\cite{Kontos02}, to a few $\units{A/cm^2}$ in Ref.~\onlinecite{Weides06b} and to about $35\units{A/cm^2}$ in the present paper, the value of $\lambda_J\propto1/\sqrt{|j_c^\pi|}$ is still above $50\units{\mu m}$. Thus, the study of a multi semifluxon system would thus require unreasonably large (mm sized) junctions.

Nonetheless, also (multifacet) junctions with length $L\lesssim\lambda_J$ are interesting. For example, one can consider an array of many alternating 0 and $\pi$ segments along $x$, where the lengths of individual segments are much smaller than $\lambda_J$. Such a structure is similar to short multifacet cuprate/Nb zigzag junctions\cite{Smilde02,Hilgenkamp03,Ariando05} or high angle grain boundaries in high $T_c$ cuprates \cite{Mannhart96}
and can \eg be used to realize a $\varphi$ junction --- a junction having a phase $\pm\varphi$ in the ground state and many other interesting properties\cite{Buzdin03,Goldobin07a,Mints98,Mints01,Mints02}.

The goal of this work is to realize Josephson junctions with various arrangements of 0 and $\pi$ segments in order to demonstrate that also complex structures are feasible.
We characterize these junctions by measurements of current voltage ($I$--$V$) characteristics, by $I_c(B)$ and by low-temperature scanning electron microscopy (LTSEM) \cite{Gross94}. By analyzing $I_c(B)$, in principle one obtains information on the suercurrent flow and (in)homogeneity of the critical current; however, the analysis at least of the more complex SIFS structures may require to consider many unknown parameters (gradients in critical current density, local inhomogeneities etc.), making conclusions ambiguous. We thus put a strong focus on LTSEM
which allows direct imaging of the supercurrent density distribution in the junctions (including counterflow areas induced by the \opi segments), close to $I_c$ \cite{Guerlich09}.

The paper is organized as follows. In Sec.~\ref{sec:samples} we discuss the sample fabrication and measurement techniques. The experimental results are presented and compared with the numerical simulations in Sec.~\ref{sec:results}. Different subsections are devoted to various geometries, (0  junction for reference, 0-$\pi$ and $0$-$\pi$-$0$ junctions, a junction consisting of 0-$\pi$ regions periodically repeated 20 times, a disk shaped structure where the 0-$\pi$ boundary forms a ring and an annular junction containing two 0-$\pi$ boundaries). All investigated samples are in the short limit ($L \lesssim 4\lambda_J$). Finally, Sec.~\ref{sec:conclusion} concludes this work.

\section{Samples and measurement techniques}
\label{sec:samples}

\subsection{Sample fabrication}

\newcommand{\oID}{\#1\xspace}
\newcommand{\pID}{\#2\xspace}
\newcommand{\opID}{\#3\xspace}
\newcommand{\opoID}{\#4\xspace}
\newcommand{\opopID}{\#5\xspace}
\newcommand{\DiskID}{\#6\xspace}
\newcommand{\RingID}{\#7\xspace}
\begin{table*}[!htb]
  \begin{tabular}{l|lccccccccc}
    \# & junction & facets & $a$ (\units{\mu m}) & $W (\units{\mu m})$ & $j_c^0 (\units{A/cm^2})$ & $|j_c^\pi| (\units{A/cm^2})$ & $\lambda_J^0$ ($\units{\mu m}$) & $\lambda_J^\pi$ ($\units{\mu m}$)& $l$ & $V_c$ ($\units{\mu V}$) \\
    \hline
    \oID    & 0                    &  1 & 50    & 10 & 85    & -     & 41   &-  & 1.2 & 50\\
    \pID    & $\pi$                &  1 & 50    & 10 & -     & 35    & -    &65 & 0.77& 18\\
    \opID   & 0-$\pi$              &  2 & 25    & 10 & 85    & 35    & 41   &65 & 1.0 & 24\\
    \opoID  & 0-$\pi$-0            &  3 & 16.6  & 10 & 73    & 33    & 44   &66 & 1.0 & 23\\
    \opopID & \OPI[20]             & 40 & 5     & 10 & 37    & 29.5  & 62   &70 & 3.0 & 11.5\\
    \DiskID &0-$\pi$ disk          & 2  & 9; 23.5& -- & 4.6   & 13.4  & 176 &103& 0.29 & 6.6\\
    \RingID &0-$\pi$ ring          & 2  & 310   & 2.5& 7.3   & 2.5   & 139  &239& 3.5 & 6.8\\
  \end{tabular}
  \caption{%
    Sample parameters at $T=4.5 \units{K}$: number of facets $N$, facet length $a$, junction width $W$. Critical current densities $j_c^0$ and $j_c^\pi$ for junctions \opID, \opoID and \opopID were estimated from fits to $I_c(B)$. $\lambda_J^0$ and $\lambda_J^\pi$ refer to the Josephson lengths of the 0 and $\pi$ parts, respectively. They are calculated from Eq.~\eqref{Eq:lambda_J} using the respective critical current densities $j_c^0$ and $j_c^\pi$. $l$ is the normalized junction length (diameter for \DiskID, circumference for \RingID), calculated from Eq.~\eqref{Eq:l}.
The characteristic voltage $V_c=I_{c}^\mathrm{max}/G$, where $G$ is the junction conductance, has been inferred by fitting the $I$--$V$ characteristic at maximum critical current $I_{c}^\mathrm{max}$ to the RSJ curve, Eq.~\eqref{Eq:RSJ}.
    For the disk shaped \ZeroPi junction the radius $r$ of the inner $\pi$ part and total (outer) radius $R$ are quoted instead of $a$. For junction \RingID the facet length $a$ is given by half of the circumference.
  }
  \label{Tab:samples}
\end{table*}

The $\mathrm{Nb|Al_2O_3|Ni_{0.6}Cu_{0.4}|Nb}$ heterostructures used for our studies were fabricated, as described in Refs.~\onlinecite{Weides06a,Weides07a}. In brief, one starts with a \(\mathrm{Nb|Al_2 O_3}\) bilayer (Nb thickness is 120\units{nm}) as for usual Nb based Josephson tunnel junctions. The thicknesses of the following F-layer must be chosen very accurately to realize 0 and $\pi$ regions with approximately the same critical current density. To achieve that, first the Ni$_{0.6}$Cu$_{0.4}$ F-layer is sputtered onto the wafer with a thickness gradient along the $y$-direction to achieve a wedge-like NiCu layer. Later on, a set of structures extending along $x$ and consisting of the 0-$\pi$ devices to be measured, plus purely 0 and $\pi$ reference junctions, is repeated several times along the $y$-direction. One of the sets will have the most suitable F-layer thickness to yield $\pi$ coupling with roughly optimal critical current density. In this way the number of wafer runs which are required to get appropriate 0-$\pi$ junctions is minimized. After the deposition of a 40 nm Nb cap-layer and lift-off one obtains a complete SIFS stack, however without steps in the thickness of the F-layer yet. To produce such steps, the parts of the structures that shall become $\pi$ regions are protected by photo resist. Then the Nb cap-layer is removed by SF$_6$ reactive rf etching, leaving a homogeneous flat NiCu surface, which is then further Ar ion etched to partially remove about 1\units{nm} of the F-layer. These areas, in the finished structures, realize the 0 regions, while the non-etched regions are $\pi$ regions. To finish the process, after removing the photo resist, a new 40\units{nm} Nb cap-layer is deposited and, after a few more photolithographic steps the full structures are completed having a 400\units{nm} thick Nb wiring layer, plus contacting leads and insulating layers. The thickness of the F-layer in the devices used here is $\sim5\units{nm}$ and is different for all devices as they come from different places of the chip because of a gradient in the F-layer thickness.

Several sets of 0, $\pi$, 0-$\pi$, 0-$\pi$-0 and \OPI[20] junctions were fabricated in the same technological run. The disk shaped and annular samples were fabricated during another run.
Parameters of the junctions are presented in Tab.~\ref{Tab:samples}.

\subsection{Measurement techniques and analysis of LTSEM signal}

For the measurements the samples were mounted on a LTSEM He cryostage and operated at a temperature $T \approx 4.5\units{K}$. Low pass filters with a cutoff frequency of 12 kHz at 4.2 K, mounted directly on the LTSEM cryostage, were used in the current and voltage leads to protect the sample from external noise. Magnetic fields of up to 1.2 mT could be applied parallel to the substrate plane and thus parallel to the junction barrier layer. We recorded $I$--$V$ characteristics and $I_c(B)$. To detect $I_c$ we used a voltage criterion $V_\mathrm{cr}$ (
0.2 $\mu \rm V$ for Figs.~\ref{fig:LTSEM_0_pi_jj} and ~\ref{fig:LTSEM_20_0_pi_jj},
0.5 $\mu \rm V$ for Figs.~\ref{fig:LTSEM_0_jj} and ~\ref{fig:LTSEM_0_pi_0_jj},
1 $\mu V$ for all other figures).

\begin{figure}[tb]
  \center{\includegraphics[width=0.5\columnwidth,clip]{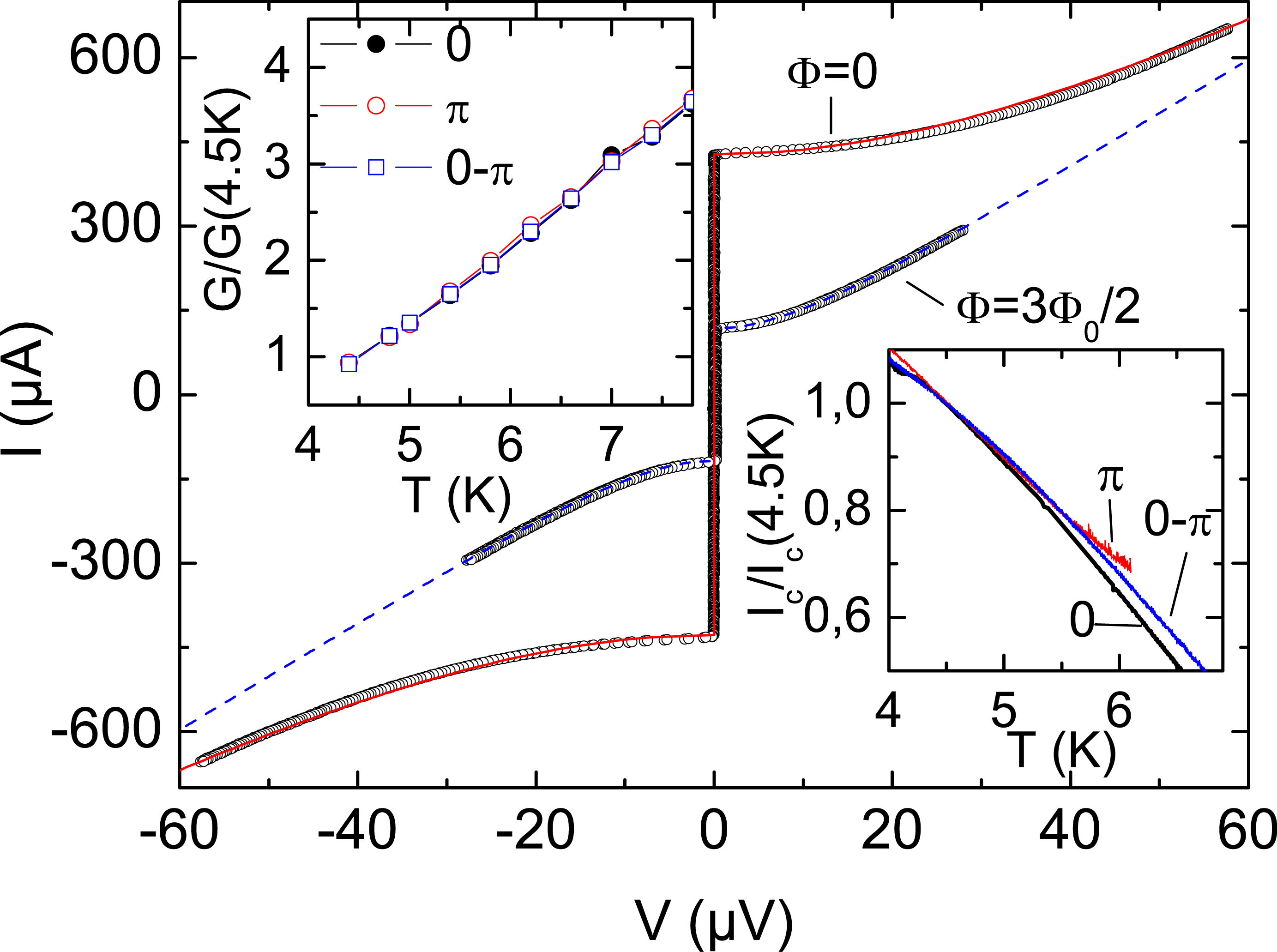}}
  \caption{(Color online)
    $I$--$V$ characteristics of a reference 0 junction (\oID in Tab.~\ref{Tab:samples}) at $T \approx 4.5\units{K}$ and applied magnetic flux $\Phi=0$ and $\nicefrac{3}{2}\Phi_0$, \ie at the principal maximum and first side maximum of $I_c(B)$. Lines correspond to the RSJ curve \eqref{Eq:RSJ}, with $I_c = 428 (118) \units{\mu A}$ and $G = 8.6 (9.7) \units{\Omega^{-1}}$ for $\Phi=0$ $(\nicefrac{3}{2}\Phi_0)$
    Upper left inset shows normalized conductances $G/G(4.5 \units{K})$ vs $T$
    for a 0, $\pi$ and \ZeroPi junction; $G(4.5K)\approx 9.5 \units{\Omega^{-1}}$ for the 0 and $0-\pi$ junction, and $\approx9.3 \units{\Omega^{-1}}$ for the $\pi$ junction.  Lower right inset shows $I_c(T)$ at $\Phi=0$, normalized to $I_c(4.5 \units{K})=420$, $170$, and $124 \units{\mu A}$ for the 0, $\pi$, and \ZeroPi junction, respectively. All junctions have dimensions of $10\times 50 \units{\mu m^2}$.
  }
  \label{fig:IVC}
\end{figure}

For selected values of magnetic field, LTSEM images were taken by recording the electron-beam-induced voltage change $\delta V(x_0,y_0)$ across the junctions (current biased slightly above $I_c$) as a function of the beam-spot coordinates $(x_0,y_0)$ on the sample surface. The periodically blanked electron beam (using $f_b \approx 6.66 \units{kHz}$, acceleration voltage $10 \units{kV}$, beam current $250 \units{pA}$), focused onto the sample, causes local heating and thus local changes in temperature-dependent parameters like the critical current density $j_c$ and conductivity $G'$ of the junction.
The beam current also adds to the bias current density in the beam spot around $(x_0,y_0)$, but for all measurements reported here the beam current density is several orders of magnitude below the typical transport current densities. Thus, this effect will be ignored here.
The local temperature rise $\delta T$ depends on the coordinates $x$, $y$ and $z$.  For our SIFS junctions the relevant depth $z_0$ is the location of the IF barrier layer, where changes in $j_c$ and $G'$ affect the $I$--$V$ characteristics by changing the critical current $I_c$ and the junction conductance $G$. We describe the temperature profile within the barrier layer of our junctions by a Gaussian distribution
\begin{equation}
  \delta T(x-x_0,y-y_0) = \Delta T\exp\left[ -\frac{(x-x_0)^2+(y-y_0)^2)}{2\sigma} \right]
  , \label{Eq:warm_spot}
\end{equation}
where  $x_0$ and $y_0$ is the position of the center of the e-beam. The LTSEM images presented below are reproduced well by simulations using $\sigma = 3.5 \units{\mu m}$; this value was used for all calculated images shown below and is somewhat larger than for other LTSEM measurements,
presumably due to the relatively thick top Nb layer. Further, from the beam-induced changes $\delta I_c$ of the critical current and the measured temperature coefficient $dI_c/dT$, we estimate $\Delta T \approx 0.5 \units{K}$. To a good approximation the beam-induced change of critical current $\delta I_c(x_0,y_0)$ is proportional to the beam-induced change of the local Josephson current density\cite{Chang84}, $\delta j_s(x_0,y_0) = j_c(x_0,y_0)\sin\phi(x_0,y_0)$ at $I_c$. To see this we write
\begin{eqnarray}
  \delta I_c &=& I_{c\rm{,on}}- I_{c\rm{,off}}
  = \int{(j_{s\rm{,on}}- j_{s\rm{,off}})df}
  \nonumber\\
  &=& \int{[j_c(T+\delta T) \sin\phi(T+\delta T)- j_c(T) \sin\phi(T)]df}.
  \nonumber\\
  &&
\end{eqnarray}
Here, the subscripts ``on'' and ``off''  refer to electron beam switched on and off. The integral $\int{(\ldots)\,df}$ has to be taken over the junction area $A_j$. The local $j_c$ depends on the coordinates $(x,y)$ via the Gaussian profile of $\delta T(x,y)$ and possible sample inhomogeneities. In addition, $j_c$ is different in the 0 and $\pi$ parts of the junction, with the values of $j_c^0$ and $j_c^\pi$ at a given temperature. Assuming that the junction is small compared to $\lambda_J$ and that a magnetic field $B$ is applied in the $(x,y)$ plane, with components $B_x$ and $B_y$ along $x$ and $y$, the Josephson phase is given by the linear ansatz
\begin{equation}
  \phi(x,y,\phi_0) = \phi_0+(2\pi/\Phi_0) \cdot \Lambda(B_y x + B_x y)
  . \label{Eq:PhaseAnsatz:Lin}
\end{equation}
At $I_c$ the initial phase $\phi_0$ is given such that the supercurrent is maximized.
For junctions having electrode thicknesses larger than the London penetration depth $\lambda_L$, the effective junction thickness is $\Lambda \approx 2\lambda_L \approx d'$. For our Nb electrodes, using $\lambda_L=90 \units{nm}$ we estimate $\Lambda \approx 180 \units{nm}$. In general, the phase $\phi$ is different in the ``on'' and ``off'' states of the beam\cite{Chang84,Chang85}. When the electron beam disturbs the junction only slightly this difference may be neglected and we obtain
\begin{equation}
  \delta I_c = \int \left[ \fracd{j_c(x,y)}{T} \cdot \sin\phi(x,y)\delta T(x-x_0,y_0) \right]\,df.
\end{equation}
As can be seen in the lower right inset of Fig.\ref{fig:IVC}, at least for some of our junctions the normalized value
\begin{equation}
  \left( \fracd{I_c}{T}\frac{1}{I_c} \right)_{B=0}=\fracd{j_c}{T}\frac{1}{j_c}
\end{equation}
(assuming a homogeneous $j_c^0, j_c^\pi$) is about constant ($\approx -0.2\units{K^{-1}}$) and roughly the same for 0 and $\pi$ parts. Note, however, that the latter statement, although valid for the junctions we study here, may not always be true. There are cases, \eg near a temperature driven 0-$\pi$ transition \cite{Ryazanov01} where $(dj_c/dT)/j_c$ of 0 and $\pi$ parts differ strongly in magnitude and perhaps even in sign. Assuming a constant value of $(dj_c/dT)/j_c$ we can further write
\begin{equation}
  \delta I_c = \fracd{j_c}{T}\frac{1}{j_c} \Delta T \img(x_0,y_0),
\label{Eq:I_c_form1}
\end{equation}
where we have used the notation
\begin{equation}
  \img(x_0,y_0) = \int \left[ j_c(x,y) \sin\phi(\phi_0,x,y)\frac{\delta T(x-x_0,y-y_0)}{\Delta T}  \right] df,
  \label{Eq:j_s_form}
\end{equation}
where the brackets indicate the convolution of $j_s$ with the beam-induced Gaussian temperature profile Eq.~\eqref{Eq:warm_spot}. When the size of the beam-induced perturbation is small compared to the structures to be imaged, we can approximate the Gaussian temperature profile with a $\delta$-function, and further simplify the above expression to

\begin{equation}
  \delta I_c \approx \fracd{j_c}{T}\frac{1}{j_c} \Delta T j_c(x_0,y_0)\sin\phi(\phi_0,x_0,y_0) A_s
  , \label{Eq:I_c_form3}
\end{equation}
with spot size $A_s \approx 2\pi\sigma^2$, defining an effective area under a 2D Gaussian distribution.
Eq.~\eqref{Eq:I_c_form3} yields $\delta I_c \propto j_s(x_0,y_0) = j_c(x_0,y_0)\sin\phi(x_0,y_0)$. Thus, by monitoring $\delta I_c$, a map of $j_s$ at $I_c$, including the supercurrent counterflow areas, can be obtained.
Note, however, that in general the spot size is not small in comparison to the structures imaged. In particular, $j_c$ sharply changes sign at a 0-$\pi$  boundary. Thus, below, we use expression \eqref{Eq:j_s_form} to calculate images  $\img(x_0,y_0)$ from the simulated supercurrent density $j_s(x,y)$ and compare them to the LTSEM images.

To obtain an LTSEM image we do not measure  $\delta I_c$ directly (the signal-to-noise ratio would be too small for reasonable measurement times which are limited by long term drifts) but bias the junctions slightly above its critical current at a given magnetic field and monitor the beam-induced voltage change $\delta V(x_0,y_0)$ as a function of the beam position $(x_0,y_0)$.
To understand in more detail the corresponding response $\delta V(x_0,y_0)$  and the experimental requirements to produce a signal proportional to $\delta I_c$ and thus proportional to $j_s$,  we first note that at the operation temperature the $I$--$V$ characteristics can be described reasonably well by the RSJ model \cite{Stewart68, Mccumber68},
\begin{equation}
  V=\sgn(I)\sqrt{I^2-I_c(B)^2}/G
\label{Eq:RSJ}
\end{equation}
for $|I|>|I_c(B)|$ and $V=0$ otherwise.
Below we will always assume $I>0$ and skip $\sgn{(I)}$.
Examples for a 0 reference junction are shown in Fig.~\ref{fig:IVC}. The $I$--$V$ characteristics have been recorded at $B=0$ and at $B=0.33\units{mT}$, corresponding to the first side maximum of $I_c(B)$. Fits to the RSJ curve are shown by lines. Note that different values of $G$ have been chosen for the two fits, which, in principle, is unphysical because $G$ should not depend on $B$. In fact, if one fits these $I$--$V$ characteristics on a large scale one would get equal values of $G$, however the region just above $I_c$ will not be approximated well, because (\ref{Eq:RSJ}) is strictly valid only for $\beta_c \equiv 2\pi I_cR^2C/\Phi_0 = 0$. In case of the $I$--$V$ characteristic for $B = 0$ we estimate that $\beta_c \sim 0.5 \ldots 0.8$.
Therefore we adopt fits with field-dependent $G$ to reproduce the $I$--$V$ characteristics near $I_c$ in the best way.

When scanning the beam over a junction, which is current-biased slightly above $I_c$, the changes $\delta I_c$ and $\delta G$ lead to a voltage change
\begin{equation}
  \delta V = -\frac{\delta G}{G^2}\sqrt{I^2-I_c(B)^2} - \frac{I_c(B)\delta I_c}{G\sqrt{I^2-I_c(B)^2}}.
  \label{Eq:Delta_V1}
\end{equation}
%
%
The change in $G$ is related to the temperature rise caused by the electron beam. Similar to the case of the critical current, $\delta G(x_0,y_0) = \int{df [(dG'/dT)\delta T(x-x_0,y-y_0)]}$. The upper left inset of Fig.\ref{fig:IVC} shows that the relative change $(dG/dT)/G = (dG'/dT)/G'$ is about constant for the junctions investigated, with a value of $0.75\units{K^{-1}}$.  We, thus, can write $\delta G = (dG'/dT)/G' \cdot \int{df [G'(x,y)\delta T(x-x_0,y-y_0)]} \approx (dG'/dT)/G' \cdot G'(x_0,y_0)\Delta T A_{s}$. In general, $G'(x_0,y_0)$ is mainly set by the insulating Al$_2$O$_3$ layer and will not strongly differ for the 0 and $\pi$ parts.
Inserting expressions for $\delta I_c$ and $\delta G$ into (\ref{Eq:Delta_V1}) we find for the beam-induced voltage change
\begin{equation}
  \delta V = \frac{I_c(B)}{G} \frac{A_{s}}{A_j} \Delta T (F_I-F_G),
  \label{Eq:Delta_V}
\end{equation}
where
%
%
\begin{equation}
    F_G = \fracd{G'}{T}\frac{1}{G'} \frac{A_jG'(x_0,y_0)}{G}
      \sqrt{[I/I_c(B)]^2-1},
  \label{Eq:F1}
\end{equation}
and
\begin{equation}
    F_I = -\fracd{j_c}{T}\frac{1}{j_c}
      \frac{A_j j_c(x_0,y_0) \sin\phi(x_0,y_0)}{I_c(B) \sqrt{[I/I_c(B)]^2-1}}.
  \label{Eq:F2}
\end{equation}
We emphasize here that these equations rely on the fact that Eq.\eqref{Eq:RSJ} provides a good fit to the $I$--$V$ characteristic in the region of interest and should at most be considered as semi-quantitative.

The response due to term $F_G$ is parasitic, if one is interested in spatial variations of the supercurrent density.  As $F_G>0$, it will give a negative and, if spatial variations of $G'(x_0,y_0)$ are small, a basically constant contribution to $\delta V$ for the whole junction area (\ie a negative offset). $F_I$ is the response of interest. To make $|F_I| \gg |F_G|$ one needs to satisfy the condition
%
\begin{equation}
  \left|\left(\fracd{G'}{T}\frac{1}{G'}\right)\left(\fracd{j_c}{T}\frac{1}{j_c}\right)^{-1}
  \frac{A_j G'(x_0,y_0)}{G} \frac{I_c(B)}{A_j j_c(x_0,y_0)\sin\phi(x_0,y_0) G}\right| \ll \frac{1}{[I/I_c(B)]^2-1}
  .\label{Eq:bias_condition_general}
\end{equation}
When the conductance is about the same for 0 and $\pi$ parts of the junction, $A_j G'(x_0,y_0)/G \approx 1$. Further, restricting requirement \eqref{Eq:bias_condition_general} to coordinates $x_0, y_0$ where $|\sin\phi(x_0,y_0)| \approx 1$ one obtains
%
\begin{equation}
  \left|\left(\fracd{G'}{T}\frac{1}{G'}\right)\left(\fracd{j_c}{T}\frac{1}{j_c}\right)^{-1}
  \frac{I_c(B)}{A_j j_c(x_0,y_0)}\right| \ll \frac{1}{[I/I_c(B)]^2-1},
  \label{Eq:bias_condition_special}
\end{equation}
with
$|(dG'/dT)(1/G')(dj_c/dT)^{-1}j_c|\approx 3.75$ for our junctions (\cf insets of Fig.\ref{fig:IVC}). As we will see, when taking images at the maxima of $I_c(B)$, at least for $A_j j_c(x_0,y_0)/I_c(B) \approx 1$, Eq.~\eqref{Eq:bias_condition_special} requires the bias current to be less than 10\% above $I_c(B)$. Note, however, that there are cases where $A_j j_c(x_0,y_0)/I_c(B)$ is large, \eg, for a homogeneous junction in high magnetic field or for a multi-facet junction when the supercurrents of the 0 and $\pi$ segments almost cancel. In this case the $F_G$ term is not dominant even much above $I_c$. On the other hand, to obtain a linear relation between $\delta V$ and $j_s(x,y)$,  $I$ should be so far above $I_c$ that $\sqrt{[I/I_c(B)]^2-1}$ varies only weakly when the beam is modulated.
Typically, this requires $I$ to be higher than about 1.05$I_c(B)$, leaving only a small window to properly bias the junction, \ie having a response $\delta V(x_0,y_0) \propto j_s(x_0,y_0)$ .

\section{Results}
\label{sec:results}

In this section we discuss $I_c(B)$ patterns and LTSEM images of a variety of SIFS junctions. All data were obtained at $T\approx 4.5\units{K}$. For reference, we will start with rectangular homogeneous 0 and $\pi$ junctions and then turn to rectangular junctions consisting of two, three and forty 0 and $\pi$ segments. Finally, we will discuss annular and disk shaped 0-$\pi$ junctions. Sketches of the different geometries are shown as insets in figures  \ref{fig:LTSEM_0_jj}(a) to \ref{fig:LTSEM_annular_jj_y}(a).

\subsection{Rectangular Junctions}
\label{sec:rect}

For all rectangular junctions of length $L$ and width $W$ we use a coordinate system with its origin at the center of the junction, so that the barrier (at $z=0$) spans from $-L/2$ to $+L/2$ in $x$ direction and from $-W/2$ to $+W/2$ in $y$ direction.

\subsubsection{0 and $\pi$ Josephson junctions}

\begin{figure}[!tb]
  \center{\includegraphics[width=0.5\columnwidth,clip]{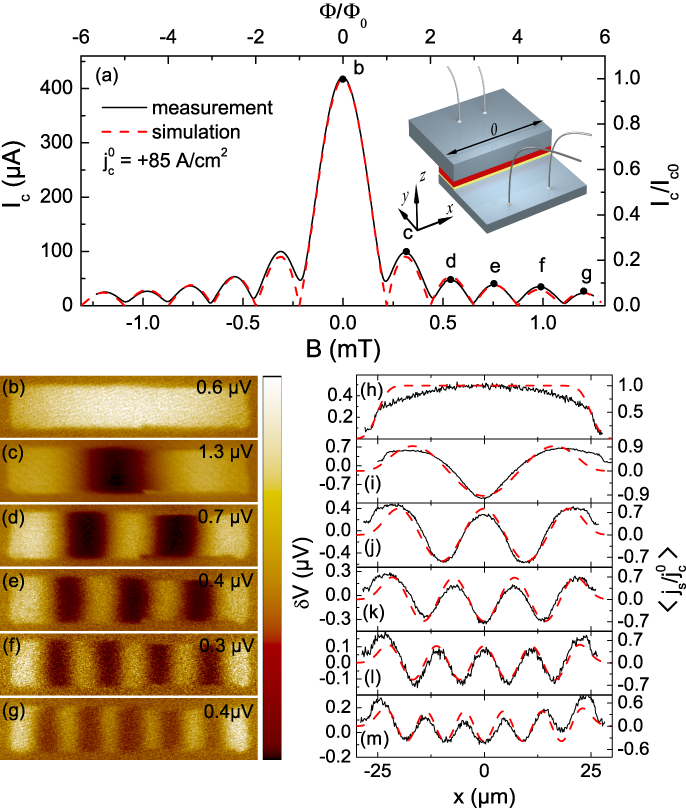}}
  \caption{(Color online). 0 junction \oID in Tab.~\ref{Tab:samples}:
    (a) $I_c(B)$ with $B \parallel y$. Solid (black) curve shows experimental data; dashed (red) curve is calculated using Eq.~\eqref{Eq:Ic(B):2D}. Inset shows the junction geometry.  (b)--(g) LTSEM images $\delta V(x,y)$\cite{FigCap} taken at bias points indicated in (a). (h)--(m) line scans: solid (black) lines $\delta V(x)$ are extracted from the corresponding LTSEM images at $y=0$; dashed (red) lines $\img(x)/j_c^0$ are calculated using a 1D version of Eq.~\eqref{Eq:j_s_form}.
  }
  \label{fig:LTSEM_0_jj}
\end{figure}

We first discuss results obtained on a 0 junction (\oID in Tab.~\ref{Tab:samples}). Fig.~\ref{fig:LTSEM_0_jj} shows $I_c(B)$ dependence, LTSEM images $\delta V(x,y)$ and corresponding line scans $\delta V(x)$ taken at $y=0$. The left hand ordinate of Fig.~\ref{fig:LTSEM_0_jj}(a) gives $I_c$ in physical units while on the right hand ordinate we have  $I_c$ normalized to $I_{c0}=A_j j_c^0$. In the graph we compare $I_c(B)$ to the Fraunhofer dependence, $I_c(B)=I_c(0)|\sin\varphi/\varphi|$, with $\varphi = \pi B \Lambda L/\Phi_0$. In fact, having more complex structures in mind, rather than using the analytic expression, we have calculated the simulated curve in Fig.~\ref{fig:LTSEM_0_jj}(a) as
\begin{equation}
  I_c(B)= \max_{\phi_0} \int_{A_j}
    \left[ j_c(x,y) \sin\phi(x,y,\phi_0) \right]\,df
  ,\label{Eq:Ic(B):2D}
\end{equation}
where $\phi(x,y,\phi_0)$ is a phase ansatz. Unless stated otherwise, we will assume a linear phase ansatz as given by Eq.~\eqref{Eq:PhaseAnsatz:Lin}. We note here that for junctions containing both 0 and $\pi$ segments $\Lambda$ may  differ by some $5\ldots10\units{\%}$ in 0 and $\pi$ regions\cite{Kemmler09, Scharinger09}. However, for the sake of simplicity, we ignore this effect here.

For the present junction we have used $j_c(x,y) = j_c^0 = const$. The resulting calculated $I_c(B)$ curve, shown by the dashed line in Fig.~\ref{fig:LTSEM_0_jj}(a), agrees with the experimental one, confirming the assumed homogeneity of $j_c^0$. From the value of $I_c(0)$ we find $j_c^0\approx 85\units{A/cm^2}$ and $\lambda_J \approx 41 \units{\mu m}$. Thus, the junction is in the short junction limit with $L/\lambda_J \approx 1.2$, justifying the use of the linear phase ansatz \eqref{Eq:PhaseAnsatz:Lin}.
Further, by comparing the abscissas of the experimental and simulated curves, one finds that $\Phi=\Phi_0$ corresponds to $B\approx 0.22 \units{mT}$. From this we obtain $\Lambda \approx 200 \units{nm}$ in good agreement with the value of $\Lambda\approx2\lambda_L\approx 180 \units{nm}$. Note that due to a magnetic field misalignment there will be a slight out-of-plane field component subject to flux focusing by large area superconducting films \cite{Ketchen85}. This leads to an increased value of $\Lambda$ calculated using the above procedure.

Fig.~\ref{fig:LTSEM_0_jj}(b) shows an LTSEM image at $B=0$. The corresponding line scan is shown by the solid line in Fig.\ref{fig:LTSEM_0_jj}(h). For $\delta V(x) \propto j_s(x) = j_c^0$ one would  expect a constant response within the junction area. The actual response is somewhat smaller at the junction edges than in the interior. Taking the finite LTSEM resolution into account, \ie calculating the convoluted supercurrent density distribution from Eq.~\eqref{Eq:j_s_form}, one obtains the dashed line which follows the measured response more closely, although there are still differences that may be caused by the junction, either by a parabolic variation of $j_c^0$ or by a variation in conductance $G'$.
To test this we implemented a parabolic variation of $j_c^0$ along $x$ in the calculation of $I_c(B)/I_c(0)$ and found that the main effect is a slight reduction of the first side minima. To still be consistent with the measured $I_c(B)$ the variation should be well below 10\% and is thus most likely not the origin of the $\delta V$ variation. To discuss a potential $G'$ effect
we quantify the $\delta V$ response using Eq.~\eqref{Eq:Delta_V}.
For the image the bias current was set to 1.05$I_c$. The function $F_G$ amounts to 0.24\units{K^{-1}} while for $F_I$ we obtain 0.62\units{K^{-1}}, \ie changes in conductance contribute by about 1/3 to the total signal.
Thus, variations of $G'$  in principle could be responsible for the observed variation of $\delta V$. However, while we could accept a simple gradient of $G'$ along $x$, the bending in $\delta V$ which is symmetric with respect to the junction center, is hard to understand.
We thus do not have a clear explanation for the parabolic shape of $\delta V(x)$.
To quantify the LTSEM response further, we can look at its maximum value $\Delta V \approx 0.45 \units{\mu V}$. With $I_c/G \approx 50 \units{\mu V}$, from Eq.~\eqref{Eq:Delta_V} one estimates $\Delta T A_{s}/A_j \approx 0.025 \units{K}$ and from that a beam-induced temperature change $\Delta T\approx 0.2 \units{K}$, which is somewhat less than $0.5\units{K}$ estimated from beam-induced $I_c$ changes.

Fig.~\ref{fig:LTSEM_0_jj}(c) shows the LTSEM image taken at the first side maximum of $I_c(B)$. The field-induced sinusoidal variation of $\delta V(x)$ can nicely be seen. The corresponding line scan is shown by the solid line in Fig.~\ref{fig:LTSEM_0_jj}(i) together with $\img(x)$, calculated using
Eq.~\eqref{Eq:j_s_form}. Here, a potential parabolic-like variation of $\delta V(x)$, if present, would be overshadowed by the stronger field-induced variation. However, the sinusoidal variation of $\delta V(x)$ with an amplitude of $0.47 \units{\mu V}$ around an offset value of $-0.13 \units{\mu V}$ points to beam-induced changes in conductance. With the bias current $I=1.1 I_c(B)$ we find $F_G \approx 0.35/\units{K}$ and  $F_I \approx -1.6/\units{K}$, \ie we expect a 20\% shift of the sinusoidal supercurrent-induced variation of $\delta V$ towards lower voltages, roughly in agreement with observation. Further, from the modulation amplitude of $0.47 \units{\mu V}$ and $I_c/G = 12 \units{\mu V}$ we estimate $\Delta T A_{s}/A_j \approx 0.025 \units{K}$ in agreement with the estimates for the zero field case.

Finally, Figs.~\ref{fig:LTSEM_0_jj}(d)--(g) show LTSEM images and Figs.~\ref{fig:LTSEM_0_jj}(j)--(m) corresponding line scans for higher order maxima in $I_c(B)$. In all cases, the field-induced modulation of $\delta V(x)$ can be seen clearly, and simulated curves for $\img(x)$, calculated using Eq.~\eqref{Eq:j_s_form}, are in good agreement with measurements.

We found similar results also for other reference junctions, including $\pi$ ones. In the latter case, typical values at $T\approx 4.5\,K$ of the critical current densities are $j_c^\pi \sim 30 \units{A/cm^2}$ (see \eg \pID in Tab.~\ref{Tab:samples}). This value is not large, but it is almost an order of magnitude higher than what has been previously reported for SIFS $\pi$ junctions\cite{Weides06b}.

\subsubsection{0-$\pi$ Josephson junction}

\begin{figure}[!tb]
  \center{\includegraphics[width=0.5\columnwidth,clip]{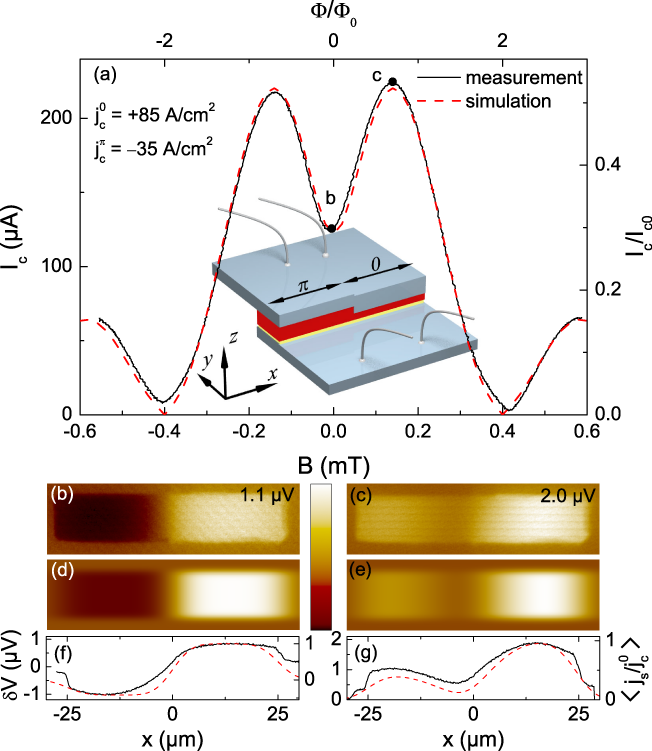}}
  \caption{(Color online). 0-$\pi$ junction \opID in Tab.~\ref{Tab:samples}:
    (a) $I_c(B)$ with $B \parallel y$. Solid (black) curve shows experimental data; dashed (red) curve is calculated using Eq.~\eqref{Eq:Ic(B):2D}. Inset shows the junction geometry. (b),(c) LTSEM images $\delta V(x,y)$\cite{FigCap} taken at bias points indicated in (a). (d),(e) corresponding images $\img(x,y)/j_c^0$ calculated using Eq.~\eqref{Eq:j_s_form}. (f),(g) line scans: solid (black) lines $\delta V(x)$ are extracted from the corresponding LTSEM images at $y=0$; dashed (red) lines $\img(x)/j_c^0$ are calculated using a 1D version of Eq.~\eqref{Eq:j_s_form}.
  }
  \label{fig:LTSEM_0_pi_jj}
\end{figure}

Now we discuss data for a 0-$\pi$ junction (\opID in Tab.~\ref{Tab:samples}) presented in Fig.~\ref{fig:LTSEM_0_pi_jj}.
The simulated $I_c(B)$ curve in Fig.~\ref{fig:LTSEM_0_pi_jj}(a) fits the experimentally measured dependence in the best way for $j_c^\pi/j_c^0=-0.42$. The right hand axis is normalized to $I_{c0}=j_c^0A_j$. From the measured value of $I_c(0)$ and the junction area $A_j$ we find $j_c^0 = 85 \units{A/cm^2}$ and  $j_c^\pi = -35 \units{A/cm^2}$. For a 0-$\pi$ junction, $\lambda_J$ can only be defined in 0 and $\pi$ parts separately, but not for the junction as a whole. However, one can find a normalized junction length as

\begin{equation}
  l \equiv \frac{L_0}{\lambda_J^0}+\frac{L_\pi}{\lambda_J^\pi}
  ,\label{Eq:l}
\end{equation}
where $L_0$ and $L_\pi$ are the total lengths of 0 and $\pi$ parts and $\lambda_J^0$ and $\lambda_J^\pi$ are the Josephson lengths in the 0 and $\pi$ parts, respectively. With this definition we calculate $l \approx 1$, showing that the junction is again in the short limit. For $\Lambda$ we obtain a reasonable value of $200 \units{nm}$.
Further note that the measured $I_c(B)$ is slightly asymmetric, \ie the main maximum at negative field is slightly lower than at positive field. This effect, which is not reproduced by the simulated curve, is due to the finite magnetization of the F-layer which, in addition, is different in the 0 and $\pi$ parts. This effect is addressed elsewhere\cite{Kemmler09}.

For the 0-$\pi$ junction, at $B$=0 the supercurrents of the two halves should have opposite sign. The part giving the smaller contribution to $I_c$ should show inverse flow of supercurrent with respect to the applied bias current, \ie, the $\pi$ part in our case. This can be seen nicely in Fig.~\ref{fig:LTSEM_0_pi_jj}(b) showing an LTSEM $\delta V(x,y)$ image at zero field. The $\pi$ part is on the left hand side. For comparison, Fig.~\ref{fig:LTSEM_0_pi_jj}(d) shows a $\img(x,y)/j_c^0$ image of the supercurrent density distribution, calculated using Eq.~\eqref{Eq:j_s_form}. For better comparison, Fig.~\ref{fig:LTSEM_0_pi_jj}(f) shows a measured and a calculated line scan. The left ordinate is shifted by $0.47 \units{\mu V}$ relative to the origin of the right ordinate to match the simulated and experimental curves. This shift is required to account for the beam-induced conductance change. More quantitatively, with $I/I_c(0) \approx 1.06$, $I_0/G=13.5 \units{\mu V}$ and assuming that $G'$ is the same for 0 and $\pi$ parts, we estimate $F_G \approx 0.3\units{K^{-1}}$. For the $\pi$ part we estimate $F_I \approx -0.8\units{K^{-1}}$, while for the 0 part we obtain $F_I \approx 1.9\units{K^{-1}}$.
The peak-to-peak voltage modulation in the LTSEM image is $1.65 \units{\mu V}$. From these numbers we estimate $(A_{s}/A_j) \Delta T \approx 0.045 \units{K}$, or $\Delta T \approx 0.3 \units{K}$, which is reasonable. For the conductance-induced shift we obtain a value of about $-0.2 \units{\mu V}$, which is about a factor of 2 less than expected from the measurement, but still within the error bars.

The LTSEM image $\delta V(x,y)$ shown in Fig.~\ref{fig:LTSEM_0_pi_jj}(c) has been taken at the main maximum of $I_c(B)$. Here, both parts of the junction give a positive response. The measurement is in good agreement with expectations, as can be seen in the calculated image $\img(x,y)/j_c^0$ in Fig.~\ref{fig:LTSEM_0_pi_jj}(e) and by comparing the line scans $\delta V(x)$ and $\img(x)/j_c^0$ shown in Fig.~\ref{fig:LTSEM_0_pi_jj}(g). Note that the ``offset problem'' seems to be less severe here. Indeed, with $I_c/G = 24 \units{\mu V}$ and $I/I_c=1.019$ we obtain $F_G \approx 0.15\units{K^{-1}}$ and $F_I \approx 2.1\units{K^{-1}}$ for the 0 part and $F_I \approx 0.85\units{K^{-1}}$ for the $\pi$ part. The supercurrent term thus clearly dominates.

\subsubsection{0-$\pi$-0 Josephson junction}

\begin{figure}[tb]
  \center{\includegraphics[width=0.5\columnwidth,clip]{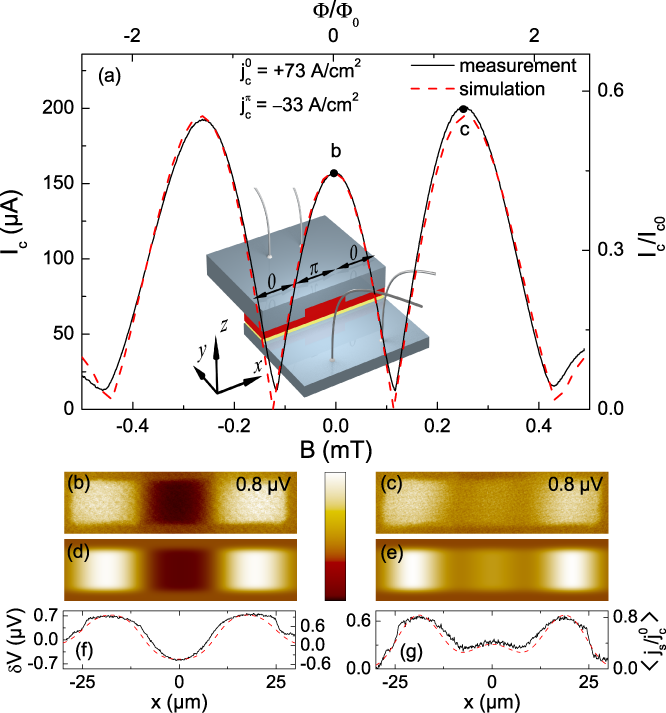}}
  \caption{(Color online). 0-$\pi$-0 junction \opoID in Tab.~\ref{Tab:samples}:
    (a) $I_c(B)$ with $B \parallel y$. Solid (black) curve shows experimental data; dashed (red) curve is calculated using Eq.~\eqref{Eq:Ic(B):2D}. Inset shows the junction geometry. (b),(c) LTSEM $\delta V(x,y)$ images\cite{FigCap} taken at bias points indicated in (a). (d),(e) corresponding images $\img(x,y)/j_c^0$ calculated using Eq.~\eqref{Eq:j_s_form}. (f),(g) line scans: solid (black) lines $\delta V(x)$ are extracted from the corresponding LTSEM images at $y=0$; dashed (red) lines $\img(x)/j_c^0$ are calculated using a 1D version of Eq.~\eqref{Eq:j_s_form}.
  }
  \label{fig:LTSEM_0_pi_0_jj}
\end{figure}

Next we discuss data for a 0-$\pi$-0 junction (\opoID in Tab.~\ref{Tab:samples}) presented in Fig.~\ref{fig:LTSEM_0_pi_0_jj}. The best fit to $I_c(B)$ was obtained for $j_c^0 = 73 \units{A/cm^2}$ and $j_c^\pi = -33 \units{A/cm^2}$. From here we obtain($l\approx 1$). We are thus again in the short junction limit. Further, we obtain $\Lambda\approx 200 \units{nm}$.

LTSEM images, taken at, respectively, the central maximum and the main maximum at positive fields, are shown in Figs.~\ref{fig:LTSEM_0_pi_0_jj}(b) and (c). Figs.~\ref{fig:LTSEM_0_pi_0_jj}(d) and (e) are simulated images, and Figs.~\ref{fig:LTSEM_0_pi_0_jj}(f) and (g) show the corresponding line scans. For this junction, the simulated curves, taking only modulations due to $j_s$ into account, agree  well with the data. For Fig.~\ref{fig:LTSEM_0_pi_0_jj}(b), with $I/I_c = 1.044$ and $I_c/G=17.5 \units{\mu V}$ we find $F_G \approx 0.22 \units{K^{-1}}$ and, for the $j_s$ maximum in the 0 part, $F_I^0 \approx 1.55\units{K^{-1}}$. For the $j_s$ maximum in the $\pi$ part we obtain $F_I^\pi \approx 0.7\units{K^{-1}}$. The offset is thus not very large. From the peak-to-peak modulation of $1.35 \units{\mu V}$ we estimate $(A_{s}/A_j) \Delta T \approx 0.035 \units{K}$ and, thus, a reasonable value $\Delta T \approx 0.23\units{K}$. Taking this value, we estimate the offset voltage to about $0.1 \units{\mu V}$.
For the measurement at the main maximum with $I/I_c=1.04$ we obtain $I_c/G = 23 \units{\mu V}$, $F_G \approx0.21\units{K^{-1}}$, $F_I^0 \approx 1.29\units{K^{-1}}$ and $F_I^\pi \approx 0.58\units{K^{-1}}$. Using $(A_{s}/A_j) \Delta T =$ 0.035 K
we expect an offset in $\delta V$ of $-0.17 \units{\mu V}$ and a maximum supercurrent response of 0.85 $\mu$V in the 0 parts, and 0.3 $\mu$V in the central $\pi$ part. The measured numbers are 0.65\units{\mu V} and 0.35\units{\mu V}, respectively.

\subsubsection{\OPI[20] Josephson junction}

\begin{figure}[tb]
 \center{\includegraphics[width=0.92\columnwidth,clip]{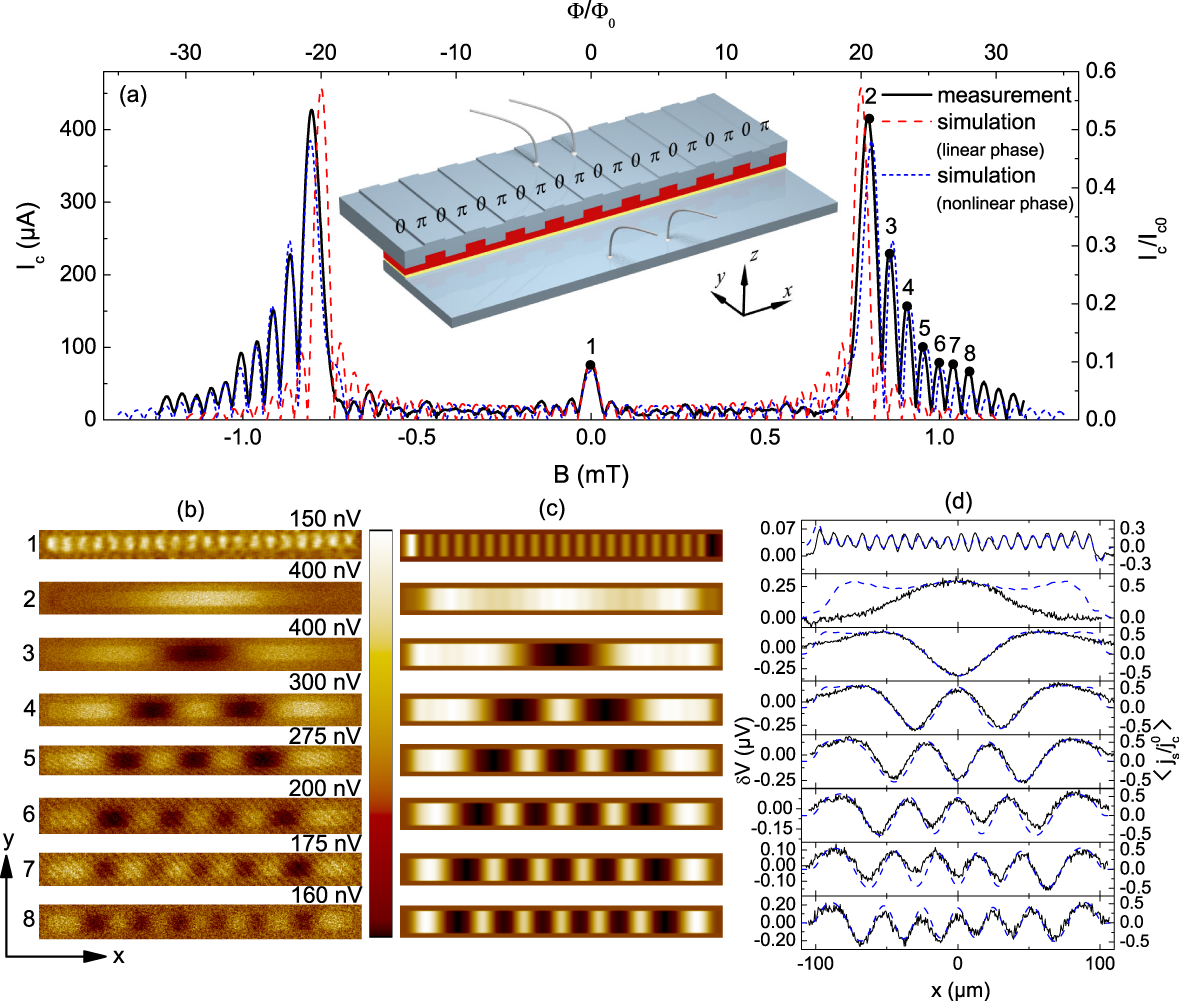}}
  \caption{(Color online). \OPI[20] junction \opopID in Tab.~\ref{Tab:samples}:
    (a) $I_c(B)$ with $B \parallel y$. Solid (black) curve shows experimental data; dashed (red) curve is calculated using Eq.~\eqref{Eq:Ic(B):2D} with linear phase ansatz \eqref{Eq:PhaseAnsatz:Lin}; dotted (blue) curve is calculated using Eq.~\eqref{Eq:Ic(B):2D} with cubic phase ansatz \eqref{Eq:PhaseAnsatz:1+3}. Inset shows the junction geometry. Only 10 \opi segments are drawn for simplicity. (b) LTSEM images $\delta V(x,y)$\cite{FigCap} taken at the bias points indicated in (a) by numbers 1 to 8. (c) corresponding images $\img(x,y)/j_c^0$ calculated using Eq.~\eqref{Eq:j_s_form} and the cubic phase ansatz \eqref{Eq:PhaseAnsatz:1+3}. (d) line scans: solid (black) lines $\delta V(x)$ are extracted from the corresponding LTSEM images at $y=0$; dashed (red) lines $\img(x)/j_c^0$ are calculated using Eq.~\eqref{Eq:j_s_form}.
  }
  \label{fig:LTSEM_20_0_pi_jj}
\end{figure}

Having seen that well behaving 0-$\pi$-0 junctions can be fabricated one may consider multisegment structures where many 0-$\pi$ segments are joined. The main purpose here is to check the complexity and reliability of the structures that can be fabricated already now.
Moreover, as already mentioned in the introduction, multi-segment \OPI[N] Josephson junctions are promising for the realization of a $\varphi$ junction.
The structure we study here has twenty \ZeroPi  segments (\opopID in Tab.~\ref{Tab:samples}). In Fig.~\ref{fig:LTSEM_20_0_pi_jj}(a) we compare the measured $I_c(B)$ dependence (solid line) with the one calculated (dashed line) using Eq.~\eqref{Eq:Ic(B):2D} with a linear phase ansatz \eqref{Eq:PhaseAnsatz:Lin}. However, on both sides of each main peak we see quite substantial deviations of the calculated curve from the experimental one. In particular, the series of $I_c$ maxima following the main peak are much higher in experiment than in simulations based on Eqs.~\eqref{Eq:Ic(B):2D} and \eqref{Eq:PhaseAnsatz:Lin}. It is interesting that such a shape of $I_c(B)$ was also measured for $d$-wave/$s$-wave zigzag shaped ramp junctions\cite{Smilde02,Ariando05,Guerlich09}.

To understand the origin of such deviations, we have tested numerically a variety of local inhomogeneities $j_c(x)$ in the different facets, ranging from random scattering to  gradients and parabolic profiles, always using the linear phase ansatz \eqref{Eq:PhaseAnsatz:Lin}. None of them, and also no variations in effective junction thickness $\Lambda(x)$ were able to qualitatively reproduce the $I_c(B)$ features described above. Finally, it turned out that the quantity to be modified is the phase ansatz, \ie, the field becomes non-uniform. Adding a cubic term, which accounts for a small phase bending, we have (assuming $B \parallel y$)
\begin{equation}
  \phi(x,y,\phi_0)=\phi_0 + 2\pi \frac{B_y \Lambda L}{2\Phi_0} \left[\frac{2x}{L}+a_3\left(\frac{2x}{L}\right)^3\right]
  . \label{Eq:PhaseAnsatz:1+3}
\end{equation}
Calculating $I_c(B)$ using Eq.~\eqref{Eq:Ic(B):2D} with $\phi$ from Eq.~\eqref{Eq:PhaseAnsatz:1+3}, we were able to reproduce the above mentioned features of the experimental $I_c(B)$ dependence, as shown by the dotted line in Fig.~\ref{fig:LTSEM_20_0_pi_jj}(a). Here we used $a_3 = -0.065$, \ie a rather small correction to the linear phase. In spite of this, for the relatively high magnetic fields around the main maxima of $I_c(B)$, this term adds up to an additional phase $\sim 1$ and becomes important --- the contribution to the integral in Eq.~\eqref{Eq:Ic(B):2D} changes essentially close to the junction ends.
Note that a homogeneous junction or a junction consisting of only a few 0 and $\pi$ segments could not sense that, since at the high fields, where the bending of the phase reaches values of $\sim 1$ at the junction edges, $I_c$ is already suppressed to almost zero.

As we will show in a separate publication \cite{Scharinger09} the origin of the nonlinear contribution in Eq.(\ref{Eq:PhaseAnsatz:1+3}) is a parasitic magnetic field component perpendicular to the junction plane, which appears due to  a misalignment $\sim 1^\circ$ between the $(x,y)$ plane and the applied magnetic field. This perpendicular component causes screening currents that result in a non-uniform (constant+parabolic) field focused inside the junction and pointing in $y$ direction. Similar effects can also be present in non-local planar junctions\cite{Moshe09}, but we are far from this limit.

By comparing the nonlinear-phase simulation to the measured $I_c(B)$ we infer $j_c^0=37 \units{A/cm^2}$, $j_c^\pi=-29.5 \units{A/cm^2}$ and $l\approx 3$. The junction is thus still in the short limit. We further obtain $\Lambda \approx 350 \units{nm}$, which is higher than the value we obtained for the other rectangular structures, but consistent with the fact that we have a focused out-of-plane field component.

Fig.~\ref{fig:LTSEM_20_0_pi_jj}(b) shows a series of LTSEM images. Image 1 is taken at $B=0$, image 2 at the main maximum and images 3 to 8 at the subsequent maxima. For image 1 one can nicely see the modulation induced by the 40 facets, although negative signals are not reached any more. This is due to the small facet size of 5 $\mu$m which is on the LTSEM resolution limit. At the main maximum the signal is strong and positive, with a slight long-range modulation but no evidence of modulations due to the individual facets any more. At the higher maxima (images 3 to 8) additional minima appear in $\delta V(x,y)$. Fig.~\ref{fig:LTSEM_20_0_pi_jj}(c) shows the corresponding images calculated using the cubic phase ansatz, and Fig.~\ref{fig:LTSEM_20_0_pi_jj}(d) shows the corresponding line scans, comparing the measured $\delta V(x)$ (solid lines) with the calculated $\img(x)$ (dotted lines). As can be seen, the agreement is excellent, except for the line scan taken at the $I_c$ maximum. Here, the measured response is strongly weakened towards the junction edges in contrast to the calculated modulation of $j_s$. For this bias, with $I=1.029I_c$ we estimate $F_G \approx 0.18\units{K^{-1}}$ and $F_I \approx 1.5\units{K^{-1}}$.  It is thus not very likely that the discrepancy is caused by a spatially varying conductance. On the other hand, from the well behaved LTSEM images at zero field we can rule out a long range variation of $j_c^0$ and $j_c^\pi$ as well. A possible origin of this behavior may be a non-uniform field focusing that results in a phase ansatz $\phi(x,y,\phi_0)$, which is more complicated than the cubic one of Eq.~\eqref{Eq:PhaseAnsatz:1+3}. However, we have to admit that we did not succeed in finding a proper dependence.

We have measured several \OPI[20] junctions. All behaved similar to the one discussed here, including the shape of $I_c(B)$ with a well developed set of maxima following the main peak and also with respect to LTSEM images. Thus, the present SIFS technology is fully able to deliberately produce quite complicated multi-facet 0-$\pi$ junctions.

\subsection{Disk Shaped and Annular Junctions}
\label{sec:disk}

\subsubsection{Disk shaped Josephson junction}

\begin{figure}[tb]
  \center{\includegraphics[width=0.5\columnwidth,clip]{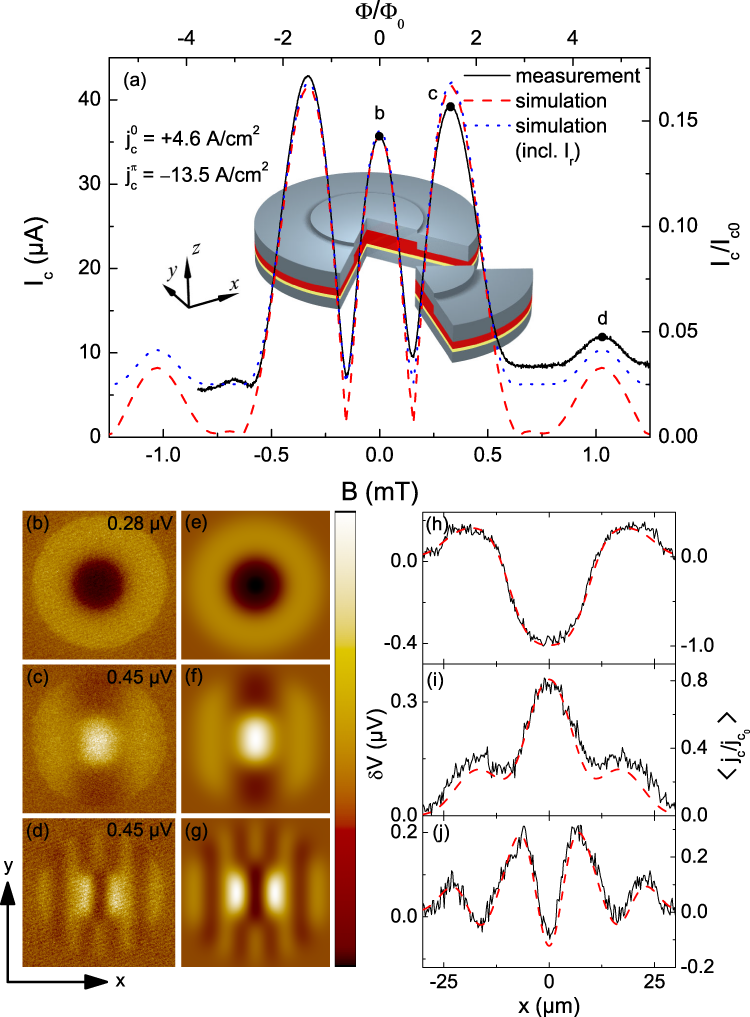}}
  \caption{(Color online). Disk shaped 0-$\pi$ junction \DiskID in Tab.~\ref{Tab:samples}:
    (a) $I_c(B)$ with $B \parallel y$. Solid (black) curve shows experimental data; dashed (red) curve is calculated using Eq.~\eqref{Eq:Ic(B):2D}; dotted (blue) curve is calculated using Eq.~\eqref{Eq:I_c^vis}. Inset shows the junction geometry. (b)--(d) LTSEM images $\delta V(x,y)$\cite{FigCap} taken at bias points indicated in (a). (e)--(g) corresponding images $\img(x,y)/j_c^0$ calculated using Eq.~\eqref{Eq:j_s_form}. (h)--(j) line scans: solid (black) lines $\delta V(x)$ are extracted from the corresponding LTSEM images at $y=0$; dashed (red) lines $\img(x)/j_c^0$ are calculated using Eq.~\eqref{Eq:j_s_form}.
  }
  \label{fig:LTSEM_disk_jj}
\end{figure}

The SIFS technology offers the possibility to create a more complex 0-$\pi$ boundary than a linear one. An intriguing option is to close this boundary in a loop. The disk shaped junction \DiskID in Tab.~\ref{Tab:samples} is of this type. Here, we use a coordinate system with its origin at the center of the disk, see the inset of Fig.~\ref{fig:LTSEM_disk_jj}(a). The $I_c(B)$ dependence, shown in Fig.~\ref{fig:LTSEM_disk_jj}(a), exhibits a central maximum at $B=0$ where the critical currents of the 0 and the $\pi$ part subtract, as well as prominent side maxima. By fitting the curve calculated using Eq.~\eqref{Eq:Ic(B):2D} (dashed line) to the experimental curve (solid line), we obtain $j_c^0 = 4.6 \units{A/cm^2}$ and $j_c^\pi = -13.4 \units{A/cm^2}$ as optimal fitting parameters. Referring to $2R$ as the junction length we obtain $l \approx 0.29$, \ie again the junction is in the short limit. Fitting the horizontal axis using the length $2R$ we obtain $\Lambda\approx 200 \units{nm}$.

For this sample, $I_c/G\approx 6.8 \units{\mu V}$ (at zero field) is rather low. As a consequence the detectability of $I_c(B)$ at low values of the critical current is resolution limited. We used a voltage criterion $V_\mathrm{cr} = 1 \units{\mu V}$ to measure the ``critical current'', yielding a parasitic $I_c$ background of $I_r=V_\mathrm{cr} G \approx 6 \units{\mu A}$. When comparing simulation with experiment the value of $I_r$ should be added (in quadrature) to the calculated critical current $I_c^\mathrm{sim}$ to obtain the ``visible critical current ''$I_c^\mathrm{vis}$, which should be compared with the experimental one $I_c^\mathrm{exp}$, \ie,
\begin{equation}
  I_c^\mathrm{vis} = \sqrt{\left( I_c^\mathrm{sim} \right)^2 + \left( I_r \right)^2}
  . \label{Eq:I_c^vis}
\end{equation}
One can see in Fig.~\ref{fig:LTSEM_disk_jj}(a) that the calculated curve including $I_r$ (dotted line) is in good agreement with the experimental data.

Fig.~\ref{fig:LTSEM_disk_jj}(b) shows an LTSEM image $\delta V(x,y)$ taken at the central maximum of $I_c(B)$. Fig.~\ref{fig:LTSEM_disk_jj}(e) shows the corresponding simulation of $\img(x,y)/j_c^0$ and Fig.~\ref{fig:LTSEM_disk_jj}(h) contains corresponding experimental and calculated line scans. The LTSEM data and the simulation results agree well, showing that the supercurrent in the central $\pi$ region flows against the bias current. Figs.~\ref{fig:LTSEM_disk_jj}(c),(f),(i) show the results for an applied magnetic field corresponding to the first side maximum of the $I_c(B)$ curve. Here, the field-induced sinusoidal variation of the supercurrent is superimposed with the disk shaped 0-$\pi$ variation. The supercurrents in the $\pi$ region as well as in a major part of the 0 region flow in the direction of the bias current, maximizing $I_c$. For completeness, in Figs.~\ref{fig:LTSEM_disk_jj}(d),(g),(j) we also show corresponding plots taken at the second side maximum of $I_c(B)$.
Here, the magnetic field induces about 7 half oscillations of the supercurrent density along $x$. Similar to the previous cases, experimental and calculated plots agree well. For the central maximum with $I/I_c = 1.09$ we find $F_G \approx 0.3\units{K^{-1}}$ and $F_I^0 \approx 2.5\units{K^{-1}}$, $F_I^\pi \approx 7.2\units{K^{-1}}$. Thus, the offset due to conductance changes is minor in this case. The same holds for the other bias points. The main reason is that the factor $j_c(x_0,y_0)A_j/I_c(B)$ entering $F_I$ is large (\eg about 7 for the $\pi$ part at $B=0$).

\subsubsection{Annular Josephson junction}

\begin{figure}[!tb]
  \center{\includegraphics[width=0.5\columnwidth,clip]{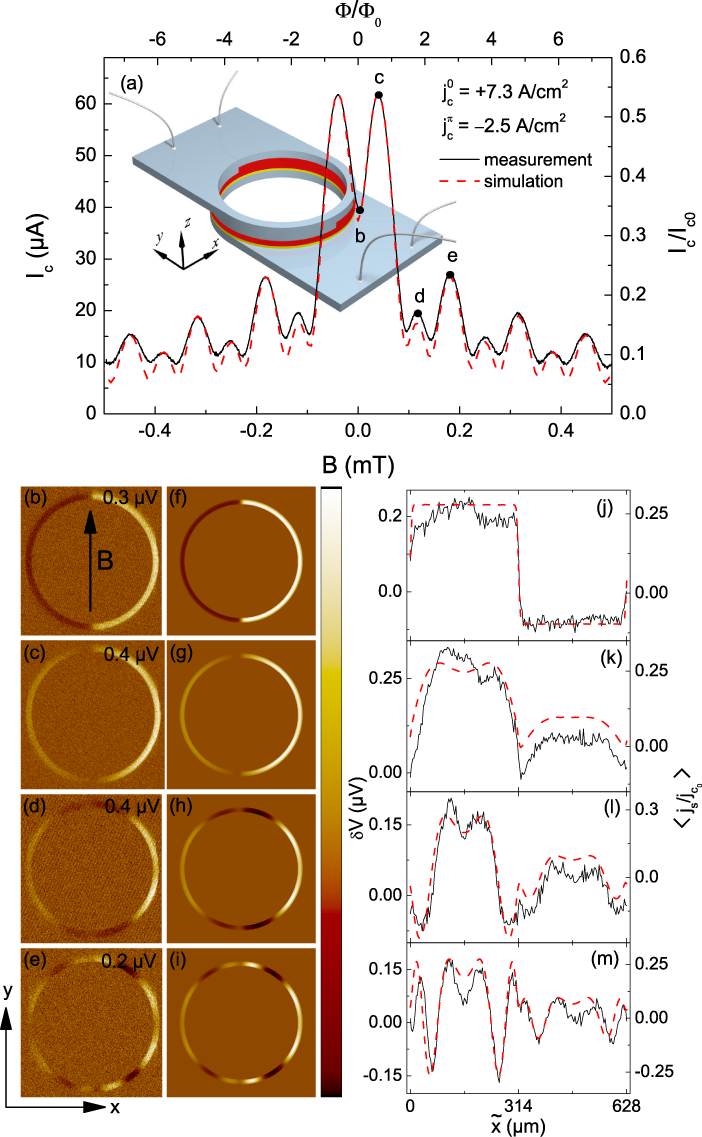}}
  \caption{(Color online). Annular 0-$\pi$ junction \RingID in Tab.~\ref{Tab:samples} with $B \parallel y$:
    (a) $I_c(B)$ pattern; solid (black) curve shows experimental data; dashed (red) curve is calculated using Eq.~\eqref{Eq:Ic(B):2D}. Inset shows the junction geometry. (b)--(e) LTSEM images\cite{FigCap} taken at bias points indicated in (a). (f)--(i) corresponding images calculated using Eq.~\eqref{Eq:j_s_form}. (g)--(m) line scans: solid (black) lines $\delta V(\tilde{x})$ are extracted from the corresponding LTSEM images; dashed (red) lines $\img(\tilde{x})/j_c^0$ are calculated using Eq.~\eqref{Eq:j_s_form} with the curvilinear coordinate $\tilde{x}$ instead of $x$ which runs along the junction circumference.
  }
  \label{fig:LTSEM_annular_jj_y}
\end{figure}

The last structure we want to discuss in this paper is an annular 0-$\pi$ junction (\RingID in Tab.~\ref{Tab:samples}, see the sketch in Fig.~\ref{fig:LTSEM_annular_jj_y}). Half of the ring is a 0 region and the other half is a $\pi$ region. One thus obtains an annular junction with two 0-$\pi$ boundaries. If the junction were long in units of  $\lambda_J$ it would be a highly interesting object to study (semi)fluxon physics, similar to the case of Nb junctions equipped with injectors\cite{Goldobin04,Kienzle09}. For this junction we use a coordinate system with its origin in the center of the ring, and the steps in the F-layer are located on the $y$ axis.  Fig.~\ref{fig:LTSEM_annular_jj_y}(a) shows $I_c(B)$ of this structure, with $B \parallel y$. The critical current is always above $10 \units{\mu A}$. This offset is in fact real and reproduced by the simulated $I_c(B)$ which is for $I_r=0$ (the actual value $I_r\approx 8 \units{\mu A}$ only slightly lifts the $I_c(B)$ minima). From the fit we obtain a ratio $j_c^\pi/j_c^0=-0.35$. Taking into account that $A_j\approx1550 \units{\mu m^2}$, we get $j_c^0 \approx 7.3 \units{A/cm^2}$ and $j_c^\pi \approx -2.5 \units{A/cm^2}$ and, referring to the circumrefence as the junction length, $l\approx 3.5$. Thus, we are still in the short junction limit. Further, we obtain $\Lambda \approx 150\units{nm}$, which is somewhat lower than for the other junctions, but still reasonable.

Figures \ref{fig:LTSEM_annular_jj_y}(b)--(e) show LTSEM images taken at various values of $B$ as labeled in the $I_c(B)$ pattern shown in Fig.~\ref{fig:LTSEM_annular_jj_y}(a). As shown in Fig.~\ref{fig:LTSEM_annular_jj_y}(b) for $B$ = 0, \ie at the central local minimum in $I_c(B)$, a counterflow in the $\pi$ part (left half) can be seen. At the main $I_c$ maximum the supercurrents in both the 0 and the $\pi$ region flow in the direction of bias current [Fig.~\ref{fig:LTSEM_annular_jj_y}(c)]. Images (d) and (e), taken at the subsequent $I_c(B)$ maxima, look more complicated, showing several regions of counterflow. In all cases, however, the LTSEM images are well reproduced by simulations, as can be seen in Figs.~\ref{fig:LTSEM_annular_jj_y}(f)--(i) and the corresponding linescans, see Figs.~\ref{fig:LTSEM_annular_jj_y}(j)--(m).
The linescans, taken along the junction circumference, start at the upper 0-$\pi$ boundary and continue clockwise.

\begin{figure}[tb]
  \center{\includegraphics[width=0.5\columnwidth,clip]{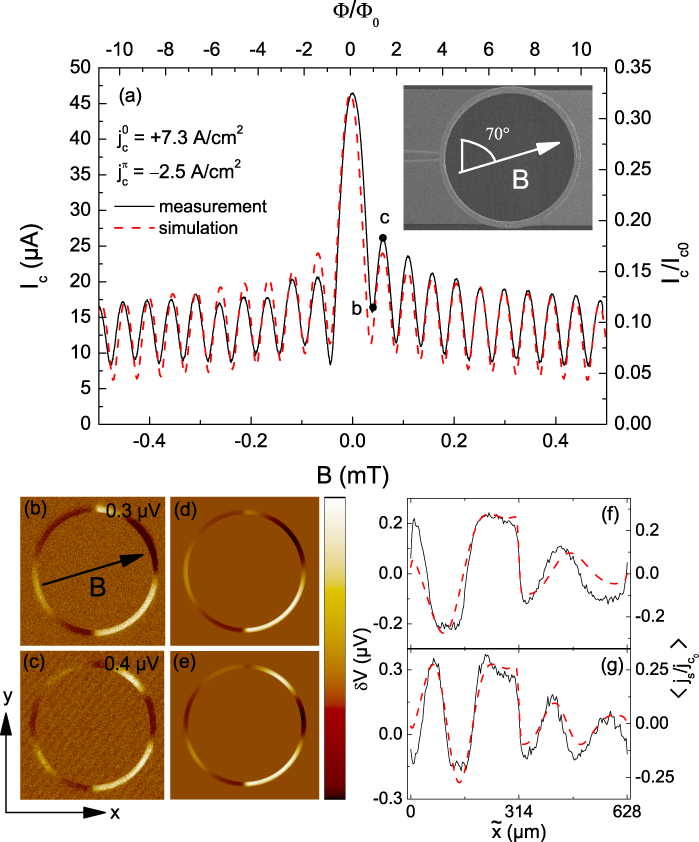}}
  \caption{(Color online). Annular 0-$\pi$ junction \RingID in Tab.~\ref{Tab:samples} with $B$ applied in the ($x,y$) plane under 70$^\circ$ from the $y$ direction:
    (a) $I_c(B)$ pattern; solid (black) curve shows experimental data; dashed (red) curve is calculated using Eq.~\eqref{Eq:Ic(B):2D}. Inset shows an SEM image of the junction. (b) and (c) LTSEM images\cite{FigCap} taken at bias points indicated in (a). (d) and (e) corresponding images calculated using Eq.~\eqref{Eq:j_s_form}. (f) and (g) line scans: solid (black) lines $\delta V(\tilde{x})$ are extracted from the corresponding LTSEM images; dashed (red) lines $\img(\tilde{x})/j_c^0$ are calculated using Eq.~\eqref{Eq:j_s_form} for $\tilde{x}$.
  }
  \label{fig:LTSEM_annular_jj_x}
\end{figure}

For this annular junction we have also rotated the magnetic field by about 70$^\circ$ towards the $x$ direction. The corresponding data are shown in Fig.~\ref{fig:LTSEM_annular_jj_x}. For this field orientation $I_c(B)$ strongly differs from the case $B \parallel y$, \cf, Fig.~\ref{fig:LTSEM_annular_jj_x}(a), but can be reproduced by simulations, using the same $j_c^0$ and $j_c^\pi$ as in Fig.~\ref{fig:LTSEM_annular_jj_y}. Furthermore, simulations show that if the field is rotated further towards the $x$ axis, the height of the side maxima in $I_c(B)$ decreases, reaching only half of their height of the 70$^\circ$ case when the field is parallel to the $x$ axis and the $I_c$ minima reach zero.
Thus, the annular 0-$\pi$ junction reacts very sensitive to field misalignments relative to the $x$ axis, similar to the case of the \OPI[20] junction where out-of-plane field components strongly altered $I_c(B)$. For completeness, Fig.~\ref{fig:LTSEM_annular_jj_x}(b)--(g) also shows LTSEM images taken at the selected bias points labeled in Fig.~\ref{fig:LTSEM_annular_jj_x}(a) and compare them with simulation. The agreement is again very good.

\section{Conclusion}
\label{sec:conclusion}

We have studied a variety of SIFS Josephson junction geometries: rectangular 0, $\pi$, $0$-$\pi$, $0$-$\pi$-$0$ and \OPI[20] junctions, disk-shaped 0-$\pi$ junction, where the 0-$\pi$ boundary forms a ring, and an annular junction with two 0-$\pi$ boundaries. Using LTSEM we were able to image the supercurrent flow in these junctions and we demonstrate that 0 and $\pi$ parts work as predicted having $j_c^0>0$ and $j_c^\pi<0$. Within each 0 or $\pi$ part, according to both LTSEM images and $I_c(B)$, the critical current density is rather homogeneous. Particularly, within our experimental resolution of a few \units{\mu m}, we saw no inhomogeneities that might have been caused by an inhomogeneous magnetization of the F-layer. This implies that ferromagnetic domains, although probably present, must have a size well below $3 \units{\mu m}$.

These results demonstrate the capabilities of the state-of-the-art SIFS technology. Arrangements like the ring-shaped 0-$\pi$ boundary are impossible to realize using other known 0-$\pi$ junction technologies\cite{Smilde02,Hilgenkamp03,Ariando05,Goldobin04}. Even intersecting  0-$\pi$ boundaries seem to be feasible, \eg, by arranging 0 and $\pi$ segments in a checkerboard pattern.

For the $\pi$ regions we demonstrated a record value of $j_c^\pi \approx 35 \units{A/cm^2}$ at $T\approx 4.5 \units{K}$, which is an order of magnitude higher than the values previously reported for SIFS junctions with a NiCu F-layer \cite{Weides06b,Weides06a}. Still, to obtain reasonable values of $\lambda_J \lesssim 20 \units{\mu m}$, $j_c^\pi$ should be increased by at least one order of magnitude to reach $\sim 1 \units{kA/cm^2}$. Then the 0-$\pi$ junctions can be made long enough (in units of $\lambda_J$) to study the dynamics of semifluxons pinned at the 0-$\pi$ boundaries. In this case semifluxon shapes, not realizable with other types of junctions, are possible, \eg, closed loops, intersecting vortices, \etc. Another issue inherent to the present SIFS technology is that the critical current densities $j_c^0$ and $j_c^\pi$ in the 0 and $\pi$ parts are not identical in general. In many cases this does not matter, \eg, when one works with semifluxons in a long junction. If $j_c^0=|j_c^\pi|$ is required, the difference in $j_c^0$ and $j_c^\pi$ will lead to a low yield of the circuit and one is perhaps restricted to operate the device in a narrow temperature interval, where $j_c^0$ and $j_c^\pi$ are closer to each other.

Even with the present restrictions, quite complex geometries like the \OPI[20] junctions have been realized. Those SIFS multifacet junctions already showed interesting features, like their high sensitivity to nonuniform magnetic fields, and they will be usable for many fundamental studies, \eg on the way of realizing $\varphi$ junctions.

\begin{acknowledgments}
  We gratefully acknowledge financial support by the Deutsche Forschungsgemeinschaft via  SFB/TRR-21 and project WE 4359/1-1, and by the German Israeli Foundation via research grant G-967-126.14/2007.
\end{acknowledgments}

\bibliography{this,SIFS_REFs}%

\begin{thebibliography}{53}
\bibitem[{Fig()}]{FigCap}
\bibinfo{note}{The color scale of all images is symmetric around zero, with the
  maximum values given on each image.}
\end{thebibliography}

\end{document}